\documentclass[11pt]{jmlr}

\usepackage{acronym}

\newacro{ML}{machine learning} 
\newacro{SSL}{self-supervised learning} 
\newacro{DL}{deep learning} 
\newacro{PCLR}{Patient Contrastive Learning of Representations} 
\newacro{CPC}{Contrastive Predictive Coding} 
\newacro{ECG}{electrocardiography}

\usepackage{longtable}

 \usepackage{booktabs}
 
\makeatletter
\def\set@curr@file#1{\def\@curr@file{#1}} 
\makeatother

\theorembodyfont{\upshape}
\theoremheaderfont{\scshape}
\theorempostheader{:}
\theoremsep{\newline}

\title[SBnCL]{Subject-based Non-contrastive Self-Supervised Learning for ECG Signal Processing}

\author{\Name{Atienza Adrian}
\Email{adar@dtu.dk}\\ 
\addr DTU Health Tech\\
DTU\\
Copenhaguen, Denmark 
\AND
\Name{Bardram Jakob}
\Email{jakba@dtu.dk}\\ 
\addr DTU Health Tech\\
DTU\\
Copenhaguen, Denmark.
\AND
\Name{Puthusserypady Sadasivan}
\Email{sapu@dtu.dk}\\ 
\addr DTU Health Tech\\
DTU\\
Copenhaguen, Denmark.}

\begin{document}

\maketitle

\begin{abstract}
Extracting information from the \acf{ECG} signal is an essential step in the design of digital health technologies in cardiology. In recent years, several \ac{ML} algorithms for automatic extraction of information in \ac{ECG} have been proposed. Supervised learning methods have successfully been used to identify specific aspects in the signal, like detection of rhythm disorders (arrhythmias). \Ac{SSL} methods, on the other hand, can be used to extract all the features contained in the data. The model is optimized without any specific goal and learns from the data itself. By adapting state-of-the-art computer vision methodologies to the signal processing domain, a few \ac{SSL} approaches have been reported recently for \ac{ECG} processing. 
However, such \ac{SSL} methods require either data augmentation or negative pairs, which limits the method to only look for similarities between two \ac{ECG} inputs, either two versions of the same signal or two signals from the same subject. This leads to models that are very effective at extracting characteristics that are stable in a subject, such as gender or age. 
But they are not successful at capturing changes within the \ac{ECG} recording that can explain dynamic aspects, like different arrhythmias or different sleep stages. 
In this work, we introduce the first \ac{SSL} method that uses neither data augmentation nor negative pairs for understanding \ac{ECG} signals, and still, achieves
 comparable quality representations. 
As a result, it is possible to design a \ac{SSL} method that not only captures similarities between two inputs, but also captures dissimilarities for a complete understanding of the data. In addition, a model based on transformer blocks is presented, which produces better results than a model based on convolutional layers (XResNet50) with almost the same number of parameters.  
\end{abstract}

\section{Introduction}

The \acf{ECG} signal contain a wealth of information not only about cardiovascular health (e.g., arrhythmias), but also for identification of  biological traits such as age and gender. In addition, the \ac{ECG} may also reveal more dynamic information about sleep stages or the person's state of mind. A comprehensive understanding and analysis of the information contained in \ac{ECG} signals is thus an important topic in \acf{ML} research for digital health.

The most widely used approach to \ac{ML} research in \ac{ECG} analysis is a supervised approach, where a cohort of annotated \ac{ECG} data is obtained and a \acf{DL} model is optimized for identification of specific features in the signal, such as heart rhythm disorder. This requires the acquisition as well as annotation of the data, which is time consuming and expensive. Furthermore, there is the challenges that the \ac{DL} model may over-fit to a specific data set, and that the model is tuned to the specific task, without having a general view of the information content of the signals. Therefore, for the same models to address similar tasks, there is a need for collection and annotation of rather large data sets. 
    
\Acf{SSL} is an emerging method in \ac{ML} and has also been applied in \ac{ECG} signal processing. It is an optimizing approach that has the potential to drive the \ac{DL} model to capture all information contained within the data and encode it into a representative vector of feature values. This vector can be used to address a wide variety of tasks afterwards, as it contains information of all kind. It relies on the capability of the model to learn useful information, directly from the data itself, rather than from the human-annotated labels associated with it. \ac{SSL} has recently been applied to \ac{ECG} processing. Methods such as \ac{PCLR}~(\cite{PCLR}) and \ac{CPC}~(\cite{12ECG}) propose different versions of \ac{SSL}. Both of these methods utilize either data augmentation, negative pairs, or both to force the model to learn informative representations. 
Data augmentation is used for obtaining two different versions of the same input. The model is driven to obtain a similar output for the two versions, by obviating the applied augmentation while focusing on the important information. In a comparable way, the use of negative pairs has the purpose of enforcing the model to compute similar representations for similar pairs, while keeping these representations different from the representations of dissimilar inputs. 
However, the use of these two techniques limits the method to only capture similarities between two \ac{ECG} inputs, which is the major drawback of these methods. As shown in this paper, these similarities are valuable but not sufficient to encode all the information contained in the data. The learned features do not represent the characteristics which are dynamic within the \ac{ECG} recording. They are the ones that are attributed with the capacity to represent heart rhythm events or sleep stages. In other words, dissimilarities also need to be codified. 
    
In this work, we demonstrate, for the first time, that there is no need of data augmentations, or negative pairs, for a \ac{DL} model to learn useful representations from the \ac{ECG} data. By avoiding using data augmentation and negative pairs, we are not restricted to just looking for similarities, meaning that it is possible to design a \ac{SSL} method which can also capture dynamic features in the \ac{ECG} signal, thereby increasing the understanding and future usage of the representation vector. In addition, we present a version of Vision Transformer as an alternative for the \ac{DL} model, which can encode richer information using the same \ac{SSL} method.


\subsection*{Generalizable Insights about Machine Learning in the Context of Healthcare}

\begin{itemize}

    \item The idea which is introduced and demonstrated in this work can be a starting point for the design of new SSL methods, specifically for ECG signals. 

    \item This method can be generalizable to other electro-physiological signals, such as electroencephalogram (EEG) or electrooculogram (EOG), since they share the same characteristics as ECG signals. They record the evolution of a vital organ over time.

    \item Convolutional Neural Networks (CNN) are the golden standard for processing vital signs. Particularly, XResNet (\cite{xresnet}) or its variants are the most used. Our work shows that Transformer models can outperform CNN-based neural network models.
\end{itemize}    

\section{Related Work}
    
In \cite{12ECG}, a novel SSL technique is designed for training a DL model. In this work, data augmentation and the use of negative pairs are used for enforcing the model to learn good representations of ECG signals. It consists on using the representation of a fixed number of sequential 10-second ECG strips for predicting the representation of a fixed number of future strips. In addition, up to 128 different versions of the signal are drawn by shuffling the order of the 10-second strips.  In a similar manner, \cite{PCLR} proposed the PCLR method that does not use data augmentation, but still uses negative pairs for the model to learn representations. This method is inspired by the study by Chen et al., (\cite{simclr}), but instead of creating two versions of the same image, as in the original paper, two 10-second ECG strips are selected from the same subject. The two strips, however, belonged to different records. The model is driven to learn similar representations when the data strips come from the same subject, while keeping these representations to be dissimilar when the data strips come from different subjects.

More recent works in computer vision have claimed that the use of negative pairs that Chen et al. (\cite{simclr}) makes use of were dispensable. Moreover, study by Grill et al. (\cite{byol}) proves that without negative pairs, a better-performance model can be optimized. This is achieved by having two versions of the same model, called student model and teacher model. In their work, two versions of the same image are given again to both the models. The student model is trained via Stochastic Gradient Descent (SGD) for predicting the representation of the teacher model, while the teacher model is an exponential moving average of the the student model. With this particular technique, the mode collapse (i.e. the model is computing the same representation,  no matter the input) is avoided without the use of negative pairs.

SSL is not the only approach for learning data representations without the use of labels. In \cite{vae}, a Variational Autoencoder (VAE) is proposed for this purpose. VAEs are a family of DL models which encode the original input into a low dimension representation which contains enough information in order to reconstruct the original input from it. Unlike the standard autoencoder, the low dimension representation is driven to belong to a known probabilistic distribution, normally Gaussian distribution. By doing this, it is possible to sample from this known distribution to create new instances. In the cited work, this latent distribution is used as the input representation.

CNN has been the most widely exploited DL architecture for processing both images as well as vital signs time series in recent years. However, in the recent times there is another DL architecture, namely, the Visual Transformer (ViT) (\cite{vit}) that achieves a comparable level of performance, if not better, than CNN based models for image processing tasks. This new architecture is based on the transformer block, which is originally described in \cite{transformer}.  

This work goes one step further than \cite{PCLR}, demonstrating that, by adapting new SSL methods in computer vision, such as \cite{byol}, not only data augmentation is dispensable, but also negative pairs. In addition, the proposed adaptation of the ViT model (\cite{vit}) for signal processing shows a higher learning capability than CNN-based models.

\section{Methods}
\subsection{Neural Network Architecture - ViT}
\begin{figure}[t]
  \centering 
  \includegraphics[width=\columnwidth]{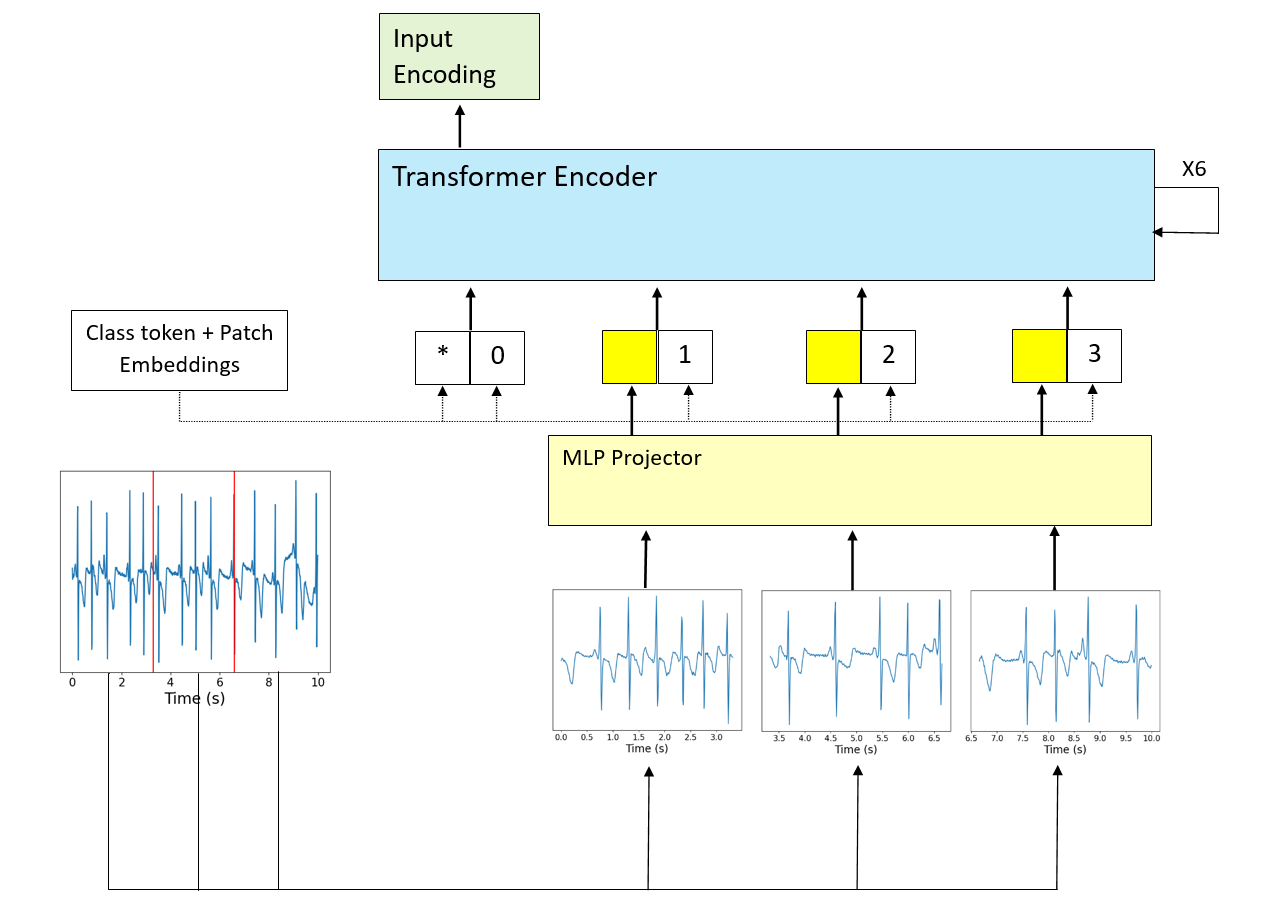} 
  \caption{Proposed DL Architecture based on ViT model }
  \label{fig:vit} 
\end{figure} 

The proposed method is inspired by the model proposed in Dosovitskiy et al. (\cite{vit}). The input size is 1000 samples (10 second signal resampled at 100 Hz) as specified in Section \ref{subsection:preprocessing}. This data strip is divided into patches of 20 values for a total of 500 patches, which are projected using a linear layer for fitting the model dimensions. The class patch is added to them and passed through a series of transformer blocks before producing the representative feature vector which is the output of the model. Figure  \ref{fig:vit} illustrates the scheme of the proposed model and the table (Table \ref{tab:hyper}) contains a complete explanation of the hyper-parameter configuration. The total number of trainable parameters is 1.192.616.
The alternative XResNet50 model used for comparing the model ability to learn has a comparable number of parameters (884.672).

\begin{table}[ht]
\centering 
\caption{Hyperparameter configuration of the proposed model.}
\vspace{0.5cm}
\begin{tabular}{lc}
\textbf{Hyper-Parameter}     & \textbf{Value}                     \\
\toprule
Model Dimension              & 128                                \\
Number of Transformer Blocks & 6                                  \\
Number of Heads              & 4                                  \\
Patch Size                   & 20                                 \\
\bottomrule
\end{tabular}
\label{tab:hyper} 
\end{table}

\subsection{Self-Supervised Learning Method - SBnCL}

\begin{figure}[t]
\centering 
\includegraphics[width=\columnwidth]{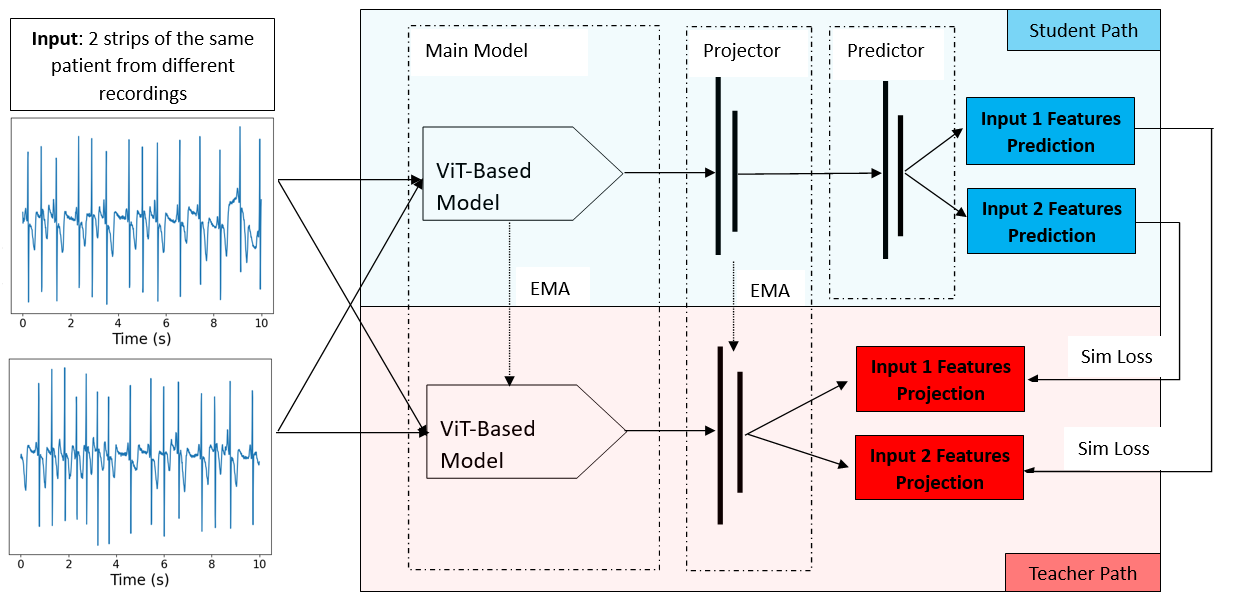} 
\caption{Schematic illustration of pretraining SSL method}
\label{fig:byol} 
\end{figure} 
The proposed SSL method is an adaptation of the method proposed by Grill et al. (\cite{byol}) in order to fit the complexity of the 10 seconds long ECG strips. This method came as an alternative of SSL contrastive learning. With the purpose of avoiding the mode collapse, an MLP projector is added in addition to the base net. A copy of this structure is used as the teacher path, which is updated as the exponential moving average of the student path. A final MLP predictor is added to the student path as well. The overall framework is updated for maximizing the capability of the student path to predict the representative vector produced by the teacher path. Figure \ref{fig:byol} illustrates the proposed framework.

As mentioned before, the teacher path is updated after every batch with the exponential moving average of the student path. The mathematical expression which capture this is the following:

\begin{equation}
\xi \leftarrow \tau \xi+(1-\tau) \theta,
\end{equation}
where $\tau$, $\xi$ and $\theta$ are the updating hyperparameter, the teacher weights and the student weights, respectively. The description of all hyperparameters proposed for this experiment are shown in Table \ref{tab:byol-hyper}. 
\begin{table}[ht]
\centering 
\caption{Hyperparameter configuration of the proposed SSL framework.}
\vspace{0.5cm}
\begin{tabular}{ll}
\textbf{SSL Hyper-Parameter}     & \textbf{Value}                     \\
\toprule
$\epsilon$                        & 1e-8                                    \\
Projector Hidden Layer Dim          & Model Dim * 2                       \\
Projector Output Layer Dim          & 64                                  \\
Predictor Hidden Layer Dim          & Model Dim * 2                       \\
Predictor Output Layer Dim          & 64                                  \\
$\tau$                              & 0.995                                  \\
\bottomrule
\end{tabular}
\label{tab:byol-hyper} 
\end{table}
The proposed loss function which measures how similar are the two representations is the following:

\begin{equation}
\text{Loss Function} = 1 - \text { Cosine similarity } = 1 - \frac{x_1 \cdot x_2}{\max \left(\left\|x_1\right\|_2 \cdot\left\|x_2\right\|_2, \epsilon\right)}, 
\end{equation}
where $x_1$ and $x_2$ are two representation vectors. $\epsilon$ has set with the value of $1e-8$

\section{Cohort}

\subsection{Cohort Selection} 

\subsubsection{Training Dataset - The Sleep Heart Health Study (SHHS)}
The open access SHHS dataset (\cite{shhs1}, \cite{shhs2}) has been used for training the model with the proposed SSL method. It consists on two cycles of polysomnography (PSG) recordings for up to 6441 and 3295 different subjects for each recording cycle, respectively. The first cycle was taken in place from 1995 to 1998, while the second one was from 2001 to 2003.

\subsubsection{Evaluation Datasets - Physionet}
The following datasets were used for evaluating the quality of the representation vector produced by the SSL trained model. They are publicly available on Physionet (\cite{physionet}).

\begin{itemize}
    \item \textbf{MIT-BIH Arrhythmia Database (MIT Arr) (\cite{mit-arr})}: This dataset contains 48 half-hour excerpts of two-channel ambulatory ECG recordings, obtained from 47 subjects from 1975 to 1979. Several heart diseases have been annotated by experts, such as Atrial Fibrillation (Afib), Atrial Flutter or Ventricular Tachycardia. 

    \item \textbf{MIT-BIH Atrial Fibrillation Database (MIT AFib) (\cite{mit-afib})}: This dataset contains 23 individual records include the two ECG signals, each one on 10 hours in duration. AFib and Atrial Flutter events have been annotated by experts. 

    \item \textbf{MIT-BIH Polysomnographic Database (MIT PSG) (\cite{mit-psg})}: This dataset contains over 80 hours of polysomnographic recordings with annotations of the sleep stages.
    
\end{itemize}

\subsection{Data Extraction} \label{subsection:preprocessing}
Since one of the aims of this work is to reproduce the results from \cite{PCLR} study, SHHS have been selected for training the SSL method. It is one of the biggest open access dataset which has two different recordings of the same subject. This is the reason that only subjects which appear in both recording cycles are used during the training procedure. This leads to 2643 subjects. ECG signals are extracted from the PSG recordings. The quality of every 10 seconds-data strips has been evaluated with the algorithm proposed by Zhao and Zhang (\cite{ecg_quality}) and just the ones with an \textit{excellent} quality are used in the training step.

In addition, signals from all specified datasets have been resampled to 100Hz and processed with a high-pass filter with cut off frequency of 0.5Hz and a filter order of 5, in order to remove the DC-offset and any baseline wander before normalization. Finally, each dataset have been normalized to have unit variance for the signals to belong to a $\mathcal{N}(0, 1)$ distribution with the purpose of removing the different device amplifications during the data collection.

\section{Results on Real Data}

\subsection{Evaluation Approach/Study Design} 
 Due to the fact that PCLR (\cite{PCLR} method is the closest to what is intended to be demonstrated, i.e., it only uses negative pairs but not data augmentation, the proposed method will be evaluated against this one. In order to evaluate the amount of information contained on the representation produced by the trained model, the following experiments have been carried out:

\begin{enumerate}
    \item The different sets on SSL method and model have been trained under same conditions (number of iterations, dataset, batch size, optimizer). In every 500 iterations, the produced features by each combination of model + SSL method have been cross-validated. Section \ref{sec:training} contains more details about this cross-validation and the results.

    \item The model is driven to extract similar features for data strips coming from the same subject. In order to evaluate this performance, a Principal Component Analysis (PCA) (\cite{pca}) is performed over the extracted features for a selected number of ECG strips. Section \ref{sec:pca_shhs} contains the details about this experiment and the results of it.

    \item In this experiment, different combinations of model + SSL method are evaluated against (\cite{vae}) for evaluating the performance of the SSL techniques compared with other non-supervised learning technique. Section \ref{sec:vs-vae} contains more details about the conducted experinment and the results of it.
    
    \item Continuing with the above-mentioned database, PCA is performed to visualize the computed features of the proposed model, with the aim of evaluating if subject-based clusters are still present with a completely unseen database. In addition, we study if the same PCA can also cluster AFib/Normal Rhythm events. Section \ref{sec:mit-arr} contains the results of these experiments.

    \item Finally, the features are evaluated for different downstream tasks, such as age regression or sleep stage classification. Results and more details of this evaluation are given in Section \ref{sec:final}
  \end{enumerate}

\subsection{Results During Training Procedure} \label{sec:training}
\begin{figure}[t]
\centering 
\includegraphics[width=0.9\columnwidth]{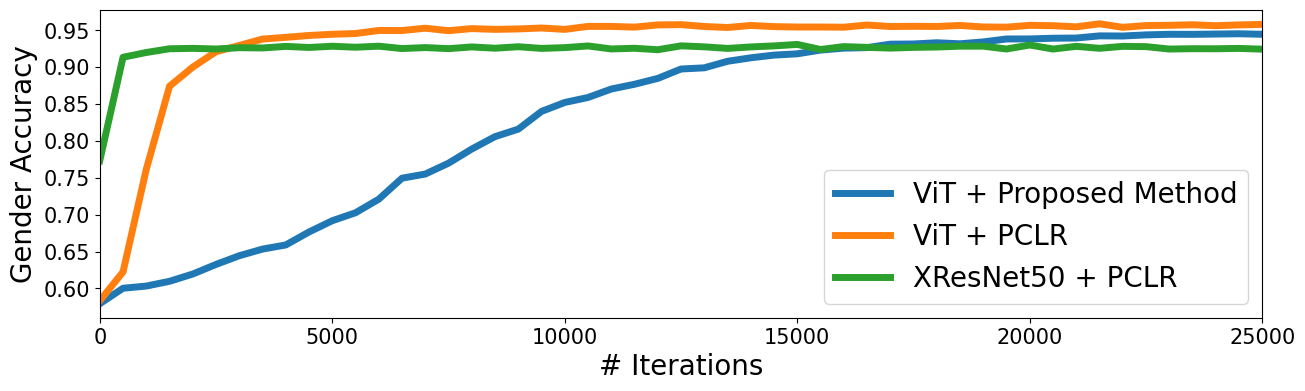} 
\caption{Gender-based cross-validation during SSL procedure described in Section \ref{sec:training}}
\label{fig:training} 
\end{figure} 
In order to track the level of information contained in the representations obtained by the model during the SSL training procedure, every 500 iterations these representation are cross-validated for the gender-classification task. For this aim, up to 15000 ECG strips have been randomly selected from the SHHS dataset. In each cross-validation, a $k$-fold procedure with 10 splits have been carried out. A K-Nearest Neighbors Algorithm (KNN) with 5 neighbours is chosen as the algorithm for performing the classification on top of the representations. Figure \ref{fig:training} shows the evolution of the chosen metric across the training procedure.

\begin{figure}[t]
\centering 
\includegraphics[width=0.85\columnwidth]{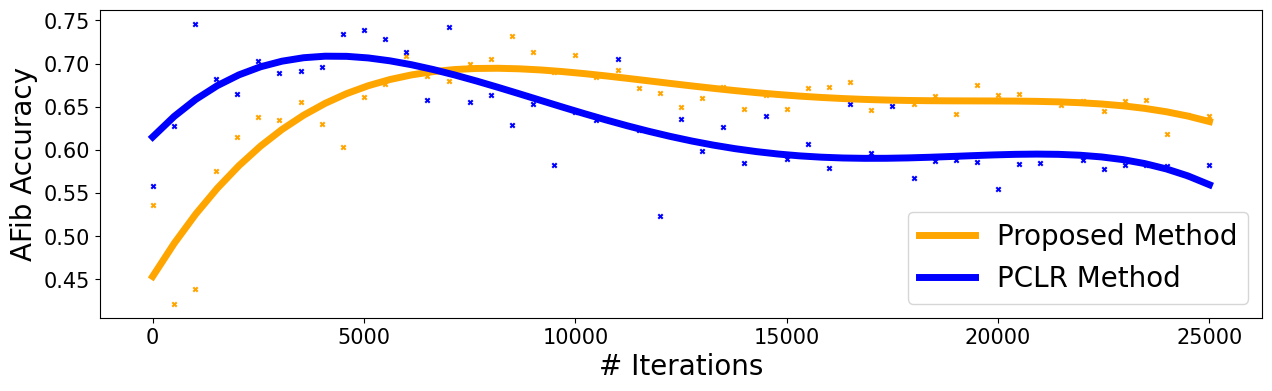} 
\caption{Cross-Evaluation of AFib accuracy for unseen subjects described in Section \ref{sec:training} across the training procedure. A polynomial has been fitted for an easier track of the evolution.}
\label{fig:afib_evol} 
\end{figure}

In a similar manner, the level of the extracted information regarding AFib events is also tracked. This experiment has been designed to test the robustness of the representations not only betweem different subjects, but also between different datasets. For this purpose, up to 768 ECG strips belonging to AFib events, in addition with 768 ECG strips belonging to normal rhythm events are selected from the MIT Arr dataset. A Support Vector Classifier (SVC) is fitted for carrying out an AFib classification task on top of the representations obtained by the model, and it is evaluated in the MIT AFib dataset. Figure \ref{fig:afib_evol} represents this evolution.

\subsection{Results on visualizing different subjects on SHHS Dataset} \label{sec:pca_shhs}
For this experiment, a subset of 11 subjects (9 of them are seen during the training process while two remains unseen) is randomly selected from the SHHS dataset. 16 ECG strips are randomly taken from these 11 subjects, half of them belonging to the first record and half of them belonging to the second, when possible (Note that unseen subjects have data from just one on the two recordings).

\begin{figure}[t]
\centering 
\includegraphics[width=0.7\columnwidth]{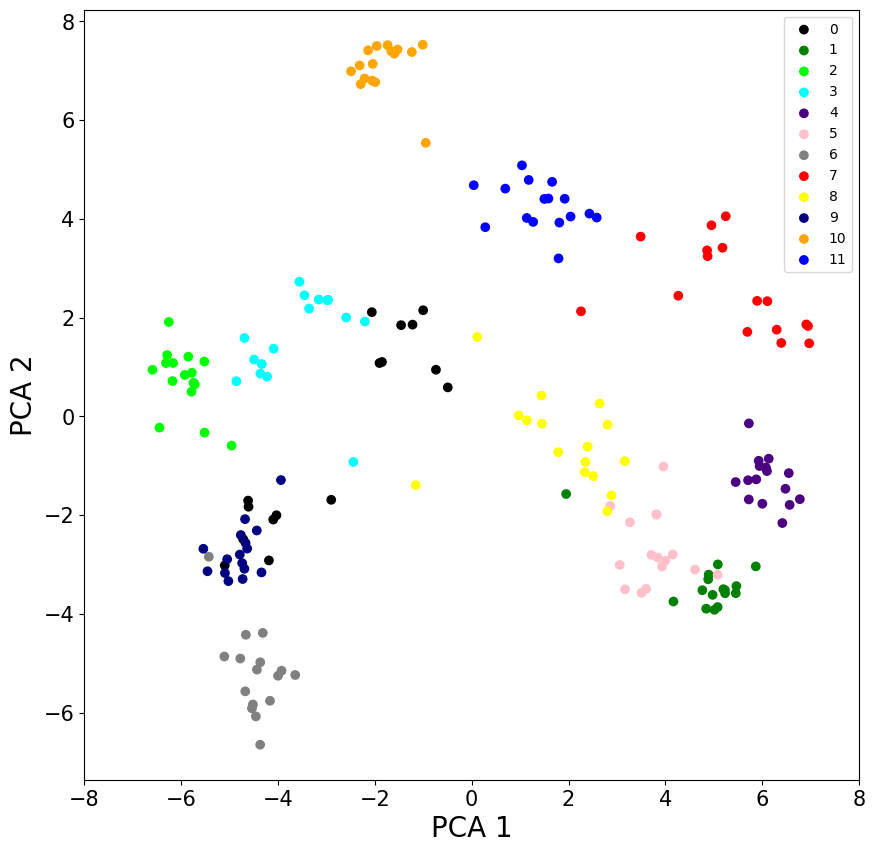} 
\caption{Subject-based PCA analysis on SHHS dataset}
\label{fig:shhs_pca_afib} 
\end{figure}

The resulting 176 different ECG strips are processed with the trained model and a PCA has been performed on compued representations. Figure \ref{fig:shhs_pca_afib} shows the different locations for each ECG strip, where subjects 10 and 11 (orange and blue dots) are the unseen subjects.

\subsection{Results on comparing VAE against SSL techniques} \label{sec:vs-vae}

In order to compare the different studies, the same dataset has been used and the same metrics have been calculated. Table \ref{tab:res-2} shows the evaluated metrics i.e., Accuracy, Sensitivity and Specificity for the three SSL/model configurations, using 1000 labeled samples, compared with VAE work reported results (In VAE study, 1118 labeled samples were used).

It is necessary to mention that in the VAE work, the model has been pre-trained on the same database on which it is evaluated afterwards. Contrary to this, all the proposed models have been pre-trained on a different database (SHHS). Furthermore, in VAE work, a neural network based classifier is trained, which is assumed to have more learning potential than the method used in our experiment, i.e., a KNN  with 5 neighbours.

\begin{table}[ht]
\centering
\caption{Results of the evaluation described in Section \ref{sec:vs-vae}}
\vspace{0.25cm}
\begin{tabular}{cccc}
Method                       & Accuracy (\%)        & Sensitivity (\%)     & Specificity (\%)     \\
\toprule
\textbf{VAE} (\cite{vae})                & 94.0           & \textbf{98.8} & 89.3         \\
\textbf{ViT-PCLR}            & \textbf{95.9} & 95.0        & 96.5         \\
\textbf{ResNET-PCLR}         & 80.6          & 74.4          & 85.2         \\
\textbf{ViT-SBnCL (Proposed)} & 95.7 & 92.7          & \textbf{97.8} \\
\bottomrule
\end{tabular}
\label{tab:res-2}
\end{table}

\subsection{Results on PCA Analyisis on AFib Dataset} \label{sec:mit-arr}
\begin{figure}[t]
\centering 
\includegraphics[width=\columnwidth]{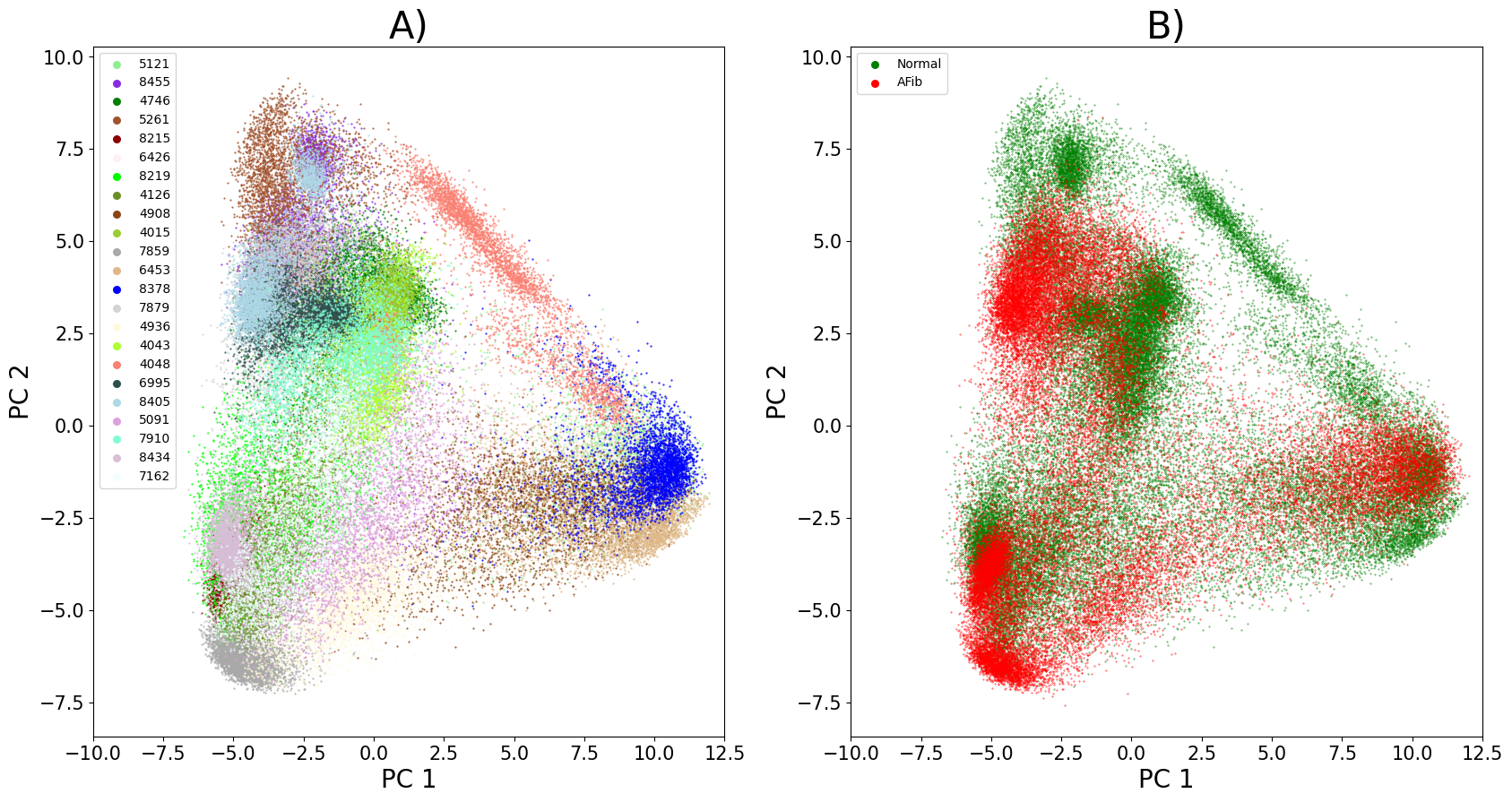} 
\caption{PCA analysis on AFib dataset: (A) Shows how the different subjects are placed in the 2D space, and (B) Captures how the AFib/Normal Rhythm events are located.}
\label{fig:pca_afib} 
\end{figure}

Using the  MIT AFib dataset, a PCA analysis has been conducted over the extracted features by the model for each ECG strip. In Figure \ref{fig:pca_afib}, it can be seen where the model is placing different strips coming from the different subjects. In addition, in the same figure, it can be seen where AFib/Normal rhythm ECG strips are localized.

\begin{figure}[t]
\centering 
\includegraphics[width=\columnwidth]{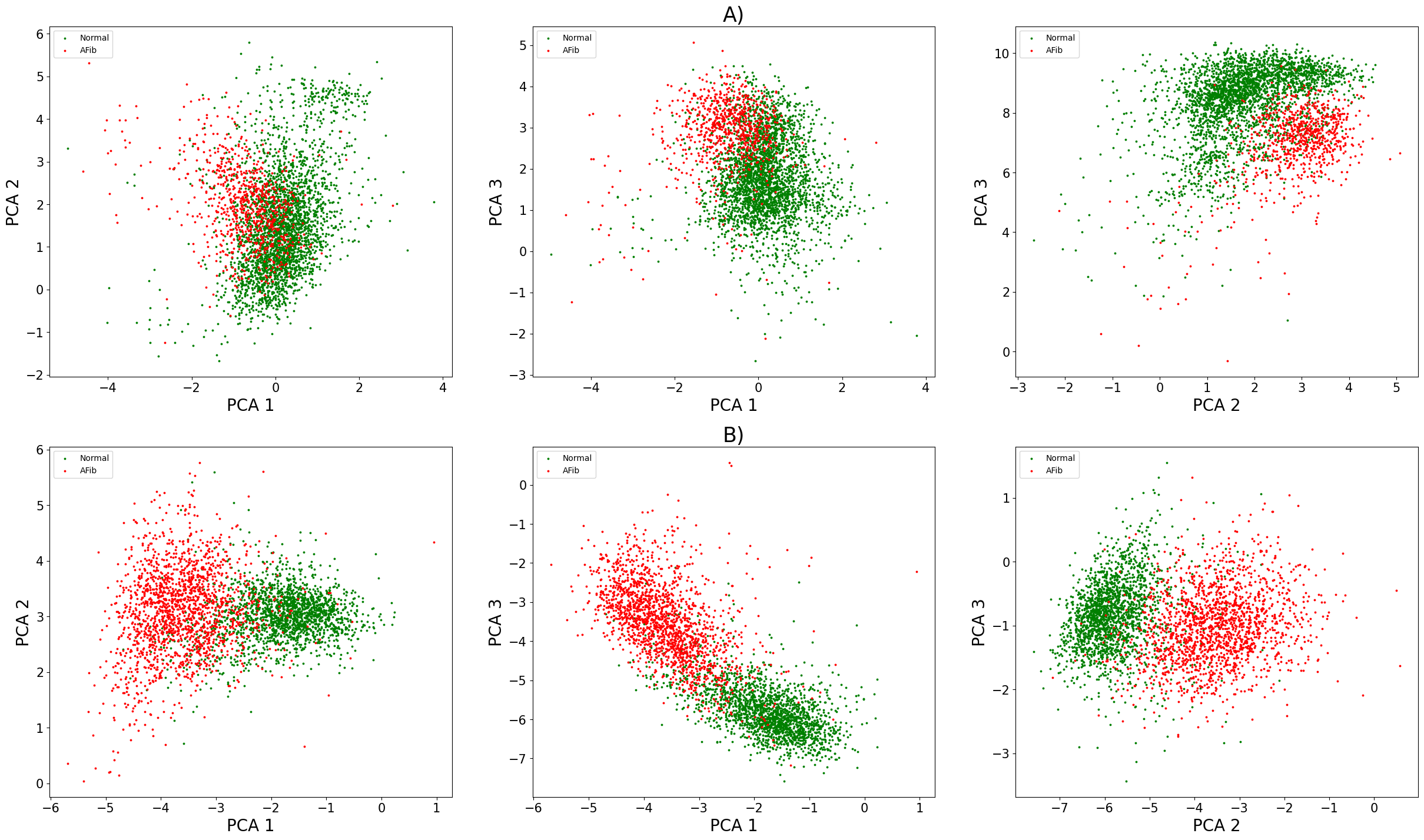} 
\caption{PCA Analysis on AFib dataset which show how AFib/Normal Rhythm events form different almost non-overlapping clusters for the selected subjects A) and B). }
\label{fig:pat_pca_afib} 
\end{figure}

For a more detailed visualization, only the PCA features belonging to two different subjects (4043 and 6995) are plotted. The Figure \ref{fig:pat_pca_afib} represents where AFib/Normal Rhythm events are placed for these specific subjects. The three different plots belong to the three different combinations of the first three components. 

\subsection{Other features evaluation} \label{sec:final} 
The purpose of this experiment is to evaluate the level of information contained on the representations for a different kind of tasks than the previously evaluated. The following two evaluations are carried out:
\begin{enumerate}
    \item The same 15000 data strips selected during the training procedure for cross-validating the gender classification task are used for predicting the age of the person. For this, a 10-fold evaluation has been run, where a linear regression model has been fitted over the output produced by the model. The metric used for evaluating its performance is the R2 score.
    \item The amount of sleep stage information that is encoded by the model is tested in this evaluation. For this purpose, MIT PSG dataset has been used. This experiment consists a 3-fold evaluation, in which 1/3 of the subjects remains as the test dataset in each K-fold iteration. A SVC is fitted on top of the model output for classifying between Awake, REM (Rapid Eye Movement) and NREM (Non-REM) stages. Accuracy is used as the metric for the evaluation.
\end{enumerate}

Table \ref{tab:res-3} contains the results of all above mentioned evaluations.

\begin{table}[t]
\centering
\caption{Evaluation of the representation vector on different tasks}
\vspace{0.25cm}
\begin{tabular}{cccc}
Method              & \multicolumn{1}{c}{Age Regression } & \multicolumn{1}{c}{Sleep Stage} \\

                  & \multicolumn{1}{c}{R2 Score} & \multicolumn{1}{c}{Accuracy (\%)} \\
\toprule
\textbf{ViT-PCLR}                                            & 0.380                                         & 53.7                                            \\
\textbf{ViT-SBnCL (Proposed)}                                    & \textbf{0.381}                                        & \textbf{55.7}         \\
\bottomrule
\label{tab:res-3}
\end{tabular}
\end{table}

\section{Discussion} 
The results obtained in this work support the hypothesis that the use of data augmentation or negative pairs techniques is dispensable in order to optimize a DL model using a SSL method. Figures \ref{fig:training} and \ref{fig:afib_evol} show that a comparable performance in gender classification and AFib classification, respectively, is achieved by the optimized model at the end of the training procedure, under the same circumstances (i.e., number of iterations, batch size, optimizer and used dataset). In addition, Table \ref{tab:res-2}  clearly indicates that a comparable performance is achieved by the model for identifying AFib/Normal rhythm events, when a few samples of each class for each subject is given to a ML algorithm. Finally, Table \ref{tab:res-3} shows that more or less same level of information about age and sleep stages is encoded by  both ViT-PCLR and ViT-SBnCL SSL methods. 

Furthermore, it has been demonstrated that the combination of SSL and Transformer-based model is the one that has the highest capacity with the aim of extracting information contained in ECG strips. While Figure \ref{fig:training} and Table \ref{tab:res-2} reflect that, under the same conditions, the use of CNN-based models clearly under-performs the proposed transformer model. Table \ref{tab:res-2} shows that SSL methods achieve better results than the VAE method, even when another dataset has been used for the model optimization.

The results show that, by capturing similarities between ECG strips from the same subject, the model is able, not only to extract information from features that are invariant throughout the recording, such as gender (as show in Figure \ref{fig:training}, but also to extract information that is unique to the subject. The model places ECG strips belonging to the same subject, even if they come from different recordings, in nearby locations in the feature map, while these locations are different from other subjects, as shown in Figure \ref{fig:shhs_pca_afib}. This ability is preserved when the model is evaluated on a different dataset than the one used for training, as shown in Figure \ref{fig:pca_afib}. In addition to this, the model not only places different subjects in different locations, but also places AFib/Normal Rhythm strips in different locations within the subject cluster as illustrated in Figure \ref{fig:pat_pca_afib}. This characteristic supports the model performance show in Table \ref{tab:res-2}, and explains that only a few observations of both events are enough for a simple ML algorithm to distinguish between AFib and normal rhythms for the rest of the data strips.

The results, however, are far from good when it comes to capturing changes in the ECG record. The representations obtained by the model do not contain any information related to AFib events that are common among subjects. This is demonstrated in Figure \ref{fig:pat_pca_afib}, where no AFib/Normal rhythm clusters are visible in the ECG strips placement, or in Figure \ref{fig:afib_evol}, where it is evident that while the model is improving in capturing similarities, the AFib classification performance decreases. Furthermore, Table \ref{tab:res-3} indicates that the representations do not explain the different sleep stages neither.

\paragraph{Conclusions:}
The proposed SSL method achieves very good performance in extracting subject-based static features/characteristics, such as age or gender from the ECG signals. However, it is unable to perform well in capturing dynamic characteristics, such as the arrhythmias or sleep stages. These features cannot be learned by the model, if the SSL method limits the model only to encode similarities. Although the results shown in this work marginally improves the results of the methods presented in other works (\cite{PCLR}), this study does open possibilities in which an SSL method can be designed for capturing dissimilarities as well, without using data augmentation and negative pairs. The model can be optimized to extract information not only about static characteristics of the subject, but also the dynamic characteristics,  such as the subject falling into different arrhythmias or different sleep stages.

\paragraph{Limitations:}
The major limitation of this work has been the dataset used to carry out the optimisation of the SSL model. To reproduce the results obtained in the PCLR study, a large dataset containing at least two recordings of the same subject, spaced in time, is necessary. The best publicly available dataset that fulfils this purpose has been the SHHS dataset. However, compared to the database used in the PCLR study, it contains a smaller number of subjects (Dataset used in PCLR study has recordings from up to 400.000 different subjects, while we are using 2643). Furthermore, this database only consists of night recordings, since they are PSG recordings. Finally, it is not possible to ensure that the used recordings contain cases of arrhythmias. Although the idea of SSL is to train a model without annotations, it would be beneficial if the database contains enough cases of all the situations for which it is intended to capture information.
\paragraph{Future Work:} Once it has been demonstrated that neither data augmentations nor negative pairs are necessary, the next logical step is to design a SSL method that captures dissimilarities as well as similarities. This model should be capable of maintaining the level of performance in stable subject characteristics and at the same time improving the amount of information contained in transient states that the person suffers throughout the recordings, such as sleep stages or different rhythm diseases.

\newpage
\bibliography{bib}

\end{document}